\documentclass[journal=jacsat,manuscript=article]{achemso}
\usepackage{amsmath}
\usepackage{amsfonts}
\usepackage{url}
\usepackage{natbib}
\title{Derivation of Casimir Effect without Zeta-Regularization}
\author{Ching-Hsuan Yen}
\email{yench@cs.nctu.edu.tw}
\affiliation{Cathay Development Center Kaohsiung, Cathay Financial Holdings, Kaohsiung 806, Taiwan}
\begin{document}
\begin{abstract}
    The Casimir effect describes the attractive force arising due to quantum fluctuations of the vacuum electromagnetic field 
between closely spaced conducting plates. 

Traditionally, zeta-regularization is employed in calculations to address infinities that emerge during the derivation. 
This paper presents a novel derivation of the Casimir effect that circumvents the need for zeta-regularization.

We derive a formula for the average vacuum energy density between two perfectly conducting plates separated by a distance. 
Our approach relates the expected vacuum energy to the change in length and position associated with each energy state. 
The uncertainty principle is incorporated to calculate the area linked to each state. 

The final result aligns with the standard Casimir effect formula obtained with zeta-regularization. 
This work demonstrates that the Casimir effect can be derived without relying on zeta-regularization, 
offering an alternative perspective on this well-established phenomenon.

\end{abstract}
\maketitle
\section{INTRODUCTION}

The Casimir effect, a cornerstone of quantum field theory, predicts an attractive force arising between closely 
spaced conducting plates due to quantum fluctuations of the vacuum electromagnetic field. 
This seemingly counterintuitive phenomenon has been experimentally verified and holds significant 
implications for miniaturized devices and nanotechnology.

Traditionally, the derivation of the Casimir effect relies on zeta-regularization, 
a mathematical technique used to handle infinities that can arise in quantum field theory calculations. 
While zeta-regularization is a powerful tool, it can introduce complexities and may obscure the underlying physical principles.

This paper presents an alternative approach to deriving the Casimir effect that avoids the use of zeta-regularization. 
Our derivation focuses on the average vacuum energy density between two perfectly conducting plates separated by a distance. 

We relate the expected vacuum energy to the change in length and position associated with each energy state confined between the plates. 
Incorporating the uncertainty principle to calculate the area linked to each energy state.
Through this approach, we arrive at a formula for the average vacuum energy density 
that aligns with the standard Casimir effect formula obtained with zeta-regularization. 
This work demonstrates that the key physical principles behind the Casimir effect can be understood without resorting to zeta-regularization, 
offering a potentially more transparent perspective on this fascinating phenomenon.

\section{Derivation}

\subsection*{Precondition}
\begin{enumerate}
    \item Uncertainty Principle\cite{heisenberg1930physical}:
    \begin{align}
        &\Delta E \Delta t \ge \frac{\hbar}{2}\\
        &\Delta x \Delta p \ge \frac{\hbar}{2}
    \end{align}
    \item The conclusion of Casimir effect assuming zeta-regularization\cite{Casimir:1948dh}: 
    \begin{align}&\frac{\langle E \rangle}{A} = -\frac{\hbar c \pi^2}{720 a^3}\end{align}
    where \(A\) is the area of the metal plates and $a$ is the distance between the metal plates.
    \item $a$ - The distance between two uncharged conductive plates in a vacuum.
    \item The plates lie parallel to the \(xy\)-plane and is orthogonal to \(z\)-axis.
\end{enumerate}

\subsection*{Calculate the Expected Vacuum Energy}

$\langle E \rangle$ represents the expected vacuum energy, 
which is the average $\bar E$ of individual energy states ($E_n$) for all possible states ($n$) in the system.
\begin{align}
    &\langle E \rangle = \sum_{n=1}^{\infty}E_n = \sum_{n=1}^{\infty}\bar{E}
\end{align}

We can express the average energy $\bar E$ as the energy uncertainty of 
a state $\Delta E_n$ divided by the ratio of its existence time $\Delta t_n$ 
to a characteristic timescale $t$.
This timescale $t$ represents the time it takes energy to travel across the gap $a$ between the plates.
\begin{align}
    &t = \frac{a}{c} \\
    &\bar{E} = \Delta E_n\frac{\Delta t_n}{t}
\end{align}

A key concept from the Heisenberg uncertainty principle states that 
the uncertainty energy of a state ($\Delta E_n$) and its corresponding uncertainty in time ($\Delta t_n$) are 
related by a constant factor ($\hbar/2$).
\begin{align}
    &\Delta E_n \Delta t_n = \frac{\hbar}{2}\\
    &\Delta E_n \frac{\Delta t_n}{t} = \frac{\hbar c }{2 a}\\
    &\langle E \rangle = \sum_{n=1}^{\infty} \frac{\hbar c }{2 a}
\end{align}
\begin{itemize}
    \item \textbf{Summation notation:} Define the expected vacuum energy (denoted by $\langle E \rangle$ ) 
    as the average of individual energy states (represented by $E_n$). 
    Since we're summing over infinitely many states, summation notation is used.
    \item \textbf{Relating energy and time:} The key idea is that the energy uncertainty ($\Delta E_n$) is related 
    to the time uncertainty ($\Delta t_n$) through a constant factor ($\hbar/2$).
    \item \textbf{Calculating energy uncertainty:} Calculate the energy uncertainty for each state ($\Delta E_n$)
     using the wavenumber and the proportionality constant from the uncertainty principle.
    \item \textbf{Deriving time uncertainty ratio:} Calculate the ratio of the time uncertainty for each state ($\Delta t_n$)
     to the total time period ($t$). This ratio is inversely proportional to the energy uncertainty.
    \item \textbf{Summing the expected energy:} This expresses the expected vacuum energy density by summing 
    the product of energy uncertainty and its corresponding time uncertainty ratio for all states, 
    divided by the total area (which will be calculated later). 
    This summation represents the average energy density across all possible quantum states.
\end{itemize}

\subsection*{Calculate the Area}

The total area $A$ is the sum of the areas $A_n$ for each individual state $n$.
The concept behind this summation is that the overall area accessible to 
a particle is determined by considering the allowed areas for each possible quantum state.
\begin{align}
    A = \sum_{n=1}^{\infty}A_n
\end{align}

    The $n$th area $A_n$ relates to the change in length and the change in position.
    \begin{align}
        A_n = L_n^2 = (\frac{a}{\Delta x_n^{x,y} \cdot n^z} \cdot a)^2
    \end{align}

    The wavenumber $k_n$ associated with the $n$th state. It relates to the momentum and wavelength of the particle in that state.
    \begin{align}
        &k_n = \frac{n \pi}{a} \\
        &p_n = \hbar k_n = \frac{n \pi \hbar}{a}
    \end{align}

    This equation calculates the change in position along the x and y directions $\Delta x_n^{x,y}$ for the $n$th state. 
It uses the Heisenberg uncertainty principle, which states that the product of momentum uncertainty and position uncertainty 
has a lower bound.
    \begin{align}
        \Delta x_n^{x, y} = \frac{\hbar/2}{p_n} = \frac{\hbar/2}{n \pi \hbar/a} = \frac{a}{2 n \pi}
    \end{align}
$n^z$ represents the ratio of a single wave in $n$th state passing through the z-axis to the plates.
    \begin{align}
        n^z = \frac{a}{2 n \pi} \cdot \frac{2 \pi}{a} =\frac{1}{n}
    \end{align}
    
Therefore

    \begin{align}
        A_n = (\frac{a}{\Delta x_n^{x,y} \cdot n^z} \cdot a)^2 = 4 n^4 \pi^2 a^2
    \end{align}
    \begin{align}
        A = \sum_{n=1}^{\infty}4 n^4 \pi^2 a^2
    \end{align}

This summation considers all possible quantum states the particle can occupy within the system 
defined by the distance between the plates.
\begin{itemize}
    \item \textbf{Relating area and uncertainty:} The concept is that the area for a particular state is related to 
    the change in length ($L_n$) and the change in position ($\Delta x_n^{x,y}$) along the x and y axes.
    \item \textbf{Defining state variables:} Define the ratio for the $n$th state ($n^z$) 
    and the momentum ($p_n$) associated with that state.
    \item \textbf{Calculating x,y position uncertainty:} Due to the confinement in the z-axis, 
    this calculates the change in position along the x and y directions for the $n$th state ($\Delta x_n^{x,y}$). 
    This is derived using the uncertainty principle and the momentum along those directions.
    \item \textbf{Expressing Area:} The formula for the area ($A_n$) of the $n$th state relates the area to the change in length, 
    change in position along x and y, and the distance between the plates.
    \end{itemize}
\subsection*{Calculate Vacuum Energy on Two Plates}
\begin{align}
    &\frac{\langle E \rangle}{A} = \sum_{n=1}^{\infty}\Delta E_n\frac{\Delta t_n}{t}\frac{1}{A_n} = \sum_{n=1}^{\infty} \frac{\hbar c }{2 a} \cdot \frac{1}{4 n^4 \pi^2 a^2} \nonumber \\
    &= \frac{\hbar c}{8 \pi^2 a^3} \sum_{n=1}^{\infty} \frac{1}{n^4} = \frac{\hbar c}{8 \pi^2 a^3} \zeta(4) = \frac{\hbar c \pi^2}{720 a^3}
\end{align}

\section{PARADOX OF CASIMIR EFFECT}

\subsection*{Precondition}
\begin{enumerate}
    \item \(L_o\) - Distance outside the plates
    \item \(L_i\) - Distance inside the plates (gap between the plates)
    \item \(P_o\) - Pressure outside the plates
    \item \(P_i\) - Pressure inside the plates
\end{enumerate}

\subsection*{Derivation}
Consider the scenario where  $L_o$ approaches infinity, while $L_i$ remains very small.
\begin{align}
    L_o \rightarrow \infty \\
    L_o \gg L_i
\end{align}

Since $L_o$ becomes infinitely large, the pressure outside plates($P_o$) remain constant regardless of the size of $L_i$.

\noindent In simpler terms:
\begin{align}L_o - Li \rightarrow L_o\end{align}

According to the Casimir effect formula\cite{Casimir:1948dh}, the pressure difference between inside and outside the plates is:
\begin{align}P_i - P_o = -\frac{\hbar c \pi^2}{240 L_i^4}\end{align}

This equation suggests that $P_o$ is greater than $P_i$

\subsection*{Situation 1}
\begin{align}
    &L_i \rightarrow 0 \\
    &P_i \ge 0 \\
    &P_o = P_i + \frac{\hbar c \pi^2}{240 L_i^4} \rightarrow {\mathbb{R}}^+ + \infty = \infty
\end{align}

This scenario leads to an unrealistic outcome. The pressure outside the plates ($P_o$) would tend towards infinity.
Consequently, the pressure difference predicted by the Casimir effect would become infinitely negative regardless of the gap size ($L_i$).
\begin{align}P_i - P_o \rightarrow P_i - \infty = - \infty\end{align}

This outcome contradicts our observations in the real world.

\subsection*{Situation 2}
\begin{align}
    &P_i < 0 \\
    &P_i \rightarrow -\frac{\hbar c \pi^2}{240 L_i^4} \\
    &P_o = P_i + \frac{\hbar c \pi^2}{240 L_i^4} \rightarrow -\frac{\hbar c \pi^2}{240 L_i^4} + \frac{\hbar c \pi^2}{240 L_i^4} = 0
\end{align}

This solution implies a negative pressure inside the platese ($P_i < 0$), 
which can be interpreted as \textbf{negative energy density} within that region.
Using the upper limit of the cosmological constant, 
the vacuum energy of free space (similar to $P_o$) has been estimated to be $5.26\times10^{-10}$(J/m$^3$)  \cite{collaboration2020planck}.
While this solution aligns better with our understanding of the universe, 
the existence of negative energy itself remains an open question and requires further investigation.

\section{CONCLUSION}

This derivation for the Casimir effect without zeta-regularization is that the average vacuum energy density 
(energy per unit area, \(\langle E \rangle/A\) ) between two perfectly conducting plates separated by distance a is:
\begin{align}\frac{\hbar c \pi^2}{720 a^3}\end{align}

This paper's derivation of the Casimir effect, lacking the negative sign,  
differs from the standard approach which attributes the negative sign to the existence of negative energy between the plates. 

While this paper calculates the absolute value of the energy density, the standard derivation focuses on the energy difference, 
where the negative sign reflects the lower energy state within the cavity due to the presence of the plates.

Since we previously established that the energy inside the plates is negative, 
the expected vacuum energy (represented by $\langle E \rangle$) is also negative.
\begin{align}
    \frac{\langle E \rangle}{A} = -\frac{\hbar c \pi^2}{720 a^3}
\end{align}

The current work has explored the theoretical implications of negative pressure within the Casimir effect. 
A crucial next step is to definitively address the question of whether negative energy truly exists.
\section{FUTURE WORK}

Definitively proving or disproving the existence of negative energy remains a captivating scientific challenge. 
The research presented here lays the groundwork for future endeavors aimed at unraveling this fundamental mystery. 
By pursuing these avenues of research, we can gain a deeper understanding of the nature 
of vacuum energy and its potential implications for our understanding of the universe.

Here are some potential avenues for future research:
\subsection*{Experimental Verification} 
    Continued advancements in experimental techniques might allow for the direct observation 
    or measurement of negative energy phenomena. 
    One potential approach could involve manipulating material properties or utilizing extreme environments 
    to create conditions conducive to negative energy states.
\subsection*{Theoretical Frameworks} 
    Further theoretical investigations could delve deeper into the nature of negative energy and 
    its compatibility with existing physical laws. 
    Exploring alternative interpretations of quantum field theory or developing new theoretical 
    frameworks might shed light on the possible existence and implications of negative energy.
\subsection*{Cosmological Implications} 
    If the existence of negative energy is confirmed, 
    its role in the broader context of cosmology would be a captivating area of inquiry. 
    Understanding how negative energy might influence phenomena like dark energy or the expansion of 
    the universe could revolutionize our understanding of the cosmos.

\bibliography{Reference}
\end{document}